\documentclass[review,12pt]{elsarticle}

\usepackage{amssymb}
\usepackage{amsthm}
\usepackage{framed} 
\usepackage{amsmath,color}
\usepackage{mathrsfs}
\usepackage{graphicx}
\usepackage{epstopdf}
\usepackage{float}
\usepackage{bm}
\usepackage{bbm}
\usepackage{mathrsfs}
\usepackage{cleveref}
\usepackage[export]{adjustbox}
\usepackage[ruled,vlined]{algorithm2e}
\usepackage{caption}
\usepackage{subcaption}
\usepackage{soul}
\usepackage{accents}
\usepackage{color,soul} 
\usepackage{color} 
\usepackage{nomencl} 
\makenomenclature 
\biboptions{sort&compress}
\soulregister\citep7 
\soulregister\citet7 
\soulregister\citealp7 
\newsavebox{\measurebox} 
\usepackage{titlesec} 
\usepackage{tabu} 
\usepackage{longtable}
\usepackage{tikz}
\usepackage{multirow}
\tikzset{>=latex}
\usepackage{pstricks}
\usepackage{listings}
\journal{Defence Technology}

\usepackage[margin=3cm]{geometry}

\newcommand{\eref}[1]{Eqn.~(\ref{#1})}

\newcommand{\fref}[1]{Fig.~\ref{#1}}

\newcommand{\tref}[1]{Table~\ref{#1}}

\newcommand{\vm}[1]{\bm{\mathrm{#1}}} 

\newcommand{\mat}[1]{\bm{\mathrm{#1}}} 

\newcommand{\cn}{\vm{n}}

\newcommand{\uu}{\mat{u}}

\makeatletter
\def\@author#1{\g@addto@macro\elsauthors{\normalsize%
    \def\baselinestretch{1}%
    \upshape\authorsep#1\unskip\textsuperscript{%
      \ifx\@fnmark\@empty\else\unskip\sep\@fnmark\let\sep=,\fi
      \ifx\@corref\@empty\else\unskip\sep\@corref\let\sep=,\fi
      }%
    \def\authorsep{\unskip,\space}%
    \global\let\@fnmark\@empty
    \global\let\@corref\@empty  
    \global\let\sep\@empty}%
    \@eadauthor={#1}
}
\makeatother

\titleformat{\paragraph}
{\normalfont\normalsize\itshape}{\theparagraph}{1em}{}
\titlespacing*{\paragraph}
{0pt}{3.25ex plus 1ex minus .2ex}{1.5ex plus .2ex}

\begin{document}

\begin{frontmatter}



\title{Adaptive phase field modelling of crack propagation in orthotropic functionally graded materials}

\author[ind]{Hirshikesh}
\author[uk]{Emilio Mart\'{\i}nez-Pa\~neda}
\author[ind]{Sundararajan Natarajan\corref{cor1}\fnref{iitm}}

\address[ind]{Department of Mechanical Engineering, Indian Institute of Technology Madras, Chennai - 600036, India.}
\address[uk]{Department of Civil and Environmental Engineering, Imperial College London, London SW7 2AZ, UK}

\cortext[cor1]{Corresponding author.}
\fntext[iitm]{Department of Mechanical Engineering, Indian Institute of Technology Madras, Chennai - 600036, India. Email: snatarajan@iitm.ac.in; sundararajan.natarajan@gmail.com}

\begin{abstract}
In this work, we extend the recently proposed adaptive phase field method to model fracture in orthotropic functionally graded materials (FGMs). A recovery type error indicator combined with quadtree decomposition is employed for adaptive mesh refinement. The proposed approach is capable of capturing the fracture process with a localized mesh refinement that provides notable gains in computational efficiency. The implementation is validated against experimental data and other numerical experiments on orthotropic materials with different material orientations. The results reveal an increase in the stiffness and the maximum force with increasing material orientation angle. The study is then extended to the analysis of orthotropic FGMs. It is observed that, if the gradation in fracture properties is neglected, the material gradient plays a secondary role, with the fracture behaviour being dominated by the orthotropy of the material. However, when the toughness increases along the crack propagation path, a substantial gain in fracture resistance is observed.
\end{abstract}



\end{frontmatter}



\newpage

\section{Introduction}
\label{Sec:Introduction}

Functionally graded materials (FGMs) are a special class of composites with spatially varying microstructure - volume fractions of the constituent elements. These characteristics of FGMs allow the designer to develop \textit{ad hoc} microstructures for specific, non-uniform service conditions. In addition, the continuous variation of material properties alleviates weak junctions within the system (for example in layered materials), i.e., avoiding the bi-material interface, which could be a potential site for crack nucleation. The potential advantages in using the FGMs include: (a) enhanced thermal and fracture resistance \cite{Aboudi1994,Pindera1998}, (b) reduced residual stresses \cite{Lee1994}, and (c) the smoothening of interfaces \cite{Ramaswamy1997,Tilbrook2006}. Ceramic-based FGMs enjoy great popularity \cite{Uemura2003}. However, these materials exhibit brittle fracture and complex fracture behaviour \cite{IJMMD2015}, particularly when a preferential direction of orthotropy develops. The preferential direction of orthotropy can arise due to the manufacturing process utilized for the synthesis. This is, for example, the case in FGMs manufactured with plasma spray techniques or electron beam physical vapor deposition. In the former, the outcome is a material with a lamellar structure with higher stiffness and weak cleavage planes parallel to the boundary. In FGMs manufactured \textit{via} electron beam physical vapor deposition one observes a columnar structure, a higher stiffness in the thickness direction and weak fracture planes perpendicular to the boundary \cite{KIM20021557,Kim2004345}.\\

Several numerical techniques have been proposed in the literature to analyse the fracture processes in orthotropic FGMs \cite{KIM20021557,HOSSEINI2013285,BAYESTEH20138,CHEN2018120,Ooi2017943,GOLI2014100}. The vast majority of the works are based on discrete approaches; for example, the conventional finite element with displacement correlation technique (DCT) \cite{Yildirim2008051106}, the extended finite element method (XFEM) \cite{KIM20021557,HOSSEINI2013285,BAYESTEH20138,GOLI2014100}, and the scaled boundary finite element method (SBFEM) \cite{CHEN2018120,Ooi2017943}. However, predicting crack initiation and subsequent crack growth requires an \textit{ad hoc} criterion, with crack trajectories being sensitive to this choice \cite{BOUCHARD2003}. Variational approaches based on energy minimization constitute a promising tool to overcome this limitation \cite{Francfort1998,Bourdin2008}. Specifically, the phase field method (PFM) has proven to be efficient technique in modelling brittle fracture \citep{McAuliffe2016,Bleyer2018,ZHOU2019729}, ductile damage \cite{Borden2016,Miehe2016b}, dynamic fracture \cite{REN201945}, fracture properties prediction of nanocomposites \cite{MSEKH2018287}, fiber cracking and composites delamination \cite{Reinoso2017a,Carollo2017, Hirshikesh2019}, plates and shells \cite{AMIRI2014102,AREIAS2016322} and hydrogen embrittlement \citep{CMAME2018,CS2020}, among other phenomena. Recently, the success of phase field fracture methods has been extended to modelling cracking in \emph{isotropic} FGMs by Hirshikesh \textit{et al.} \cite{HIRSHIKESH2019239}. Here, we extend the framework to deal with orthotropic FGMs and include an adaptive mesh refinement strategy to boost computational efficiency.\\

Although, the PFM has shown advantages over discrete approaches, the finite element discretization requires resolving the the length scale parameter as $\ell_o$. In brittle materials, $\ell_o$ can be very small and the discrete crack in linear elastic fracture mechanics is recovered for the limiting case of $\ell_o \rightarrow 0$. The need to resolve this region of high gradients creates a computational burden. Local refinement techniques can reduce the computational cost; however, this requires the crack path to be known \textit{a priori}, which is often not the case. An alternative is to use adaptive refinement algorithms based on error indicators. Several strategies have been proposed \citep{PATIL2018254,Tian20191108,Mang2003,HIRSHIKESH2019106599,Mahnken2013418,HIRSHIKESH2019284}, being most of them based on post-error estimation such as goal-oriented, recovery, and residual. For example, Areias \textit{et al.} \cite{AREIAS2018339,AREIAS2016116} presented an adaptive mesh refinement strategy that combines the staggered algorithm with the screened Poisson equation. Goswami \textit{et al.} \cite{Goswami2020112808} proposed an adaptive fourth-order phase field model based isogeometric analysis (IGA). Recently, Samaniego \textit{et al.}  \cite{Samaniego2020112790} solved the phase-field equations via machine learning approach. In this paper, we aim to extend the recently developed adaptive PFM by Hirshikesh \textit{et al.} \cite{HIRSHIKESH2019106599} to model fracture in orthotropic FGMs. The adaptive PFM is based on the combination of quadtree decomposition and recovery based error indicators, allowing for an automatic tracking of the crack trajectory and local domain discretization. The hanging nodes that arise due to the quadtree decomposition are treated within the framework of the polygonal finite element method (PFEM) with mean-value coordinate basis function.\\  

The rest of the paper is organized as follows. Section \ref{Sec:NumModel} presents the governing equations for the PFM and the corresponding weak form. The adaptive refinement strategy based on the quadtree decomposition and recovery based error indicator is presented in Section \ref{Sec:Quadtree Meshing}. The applicability of the adaptive refinement strategy for the fracture in orthotropic FGM is shown in Section \ref{Sec:Results}. Concluding remarks end the manuscript.

\section{A phase field fracture formulation for orthotropic FGMs}
\label{Sec:NumModel}
\begin{figure}[htpb]
\centering
\includegraphics[scale=0.5]{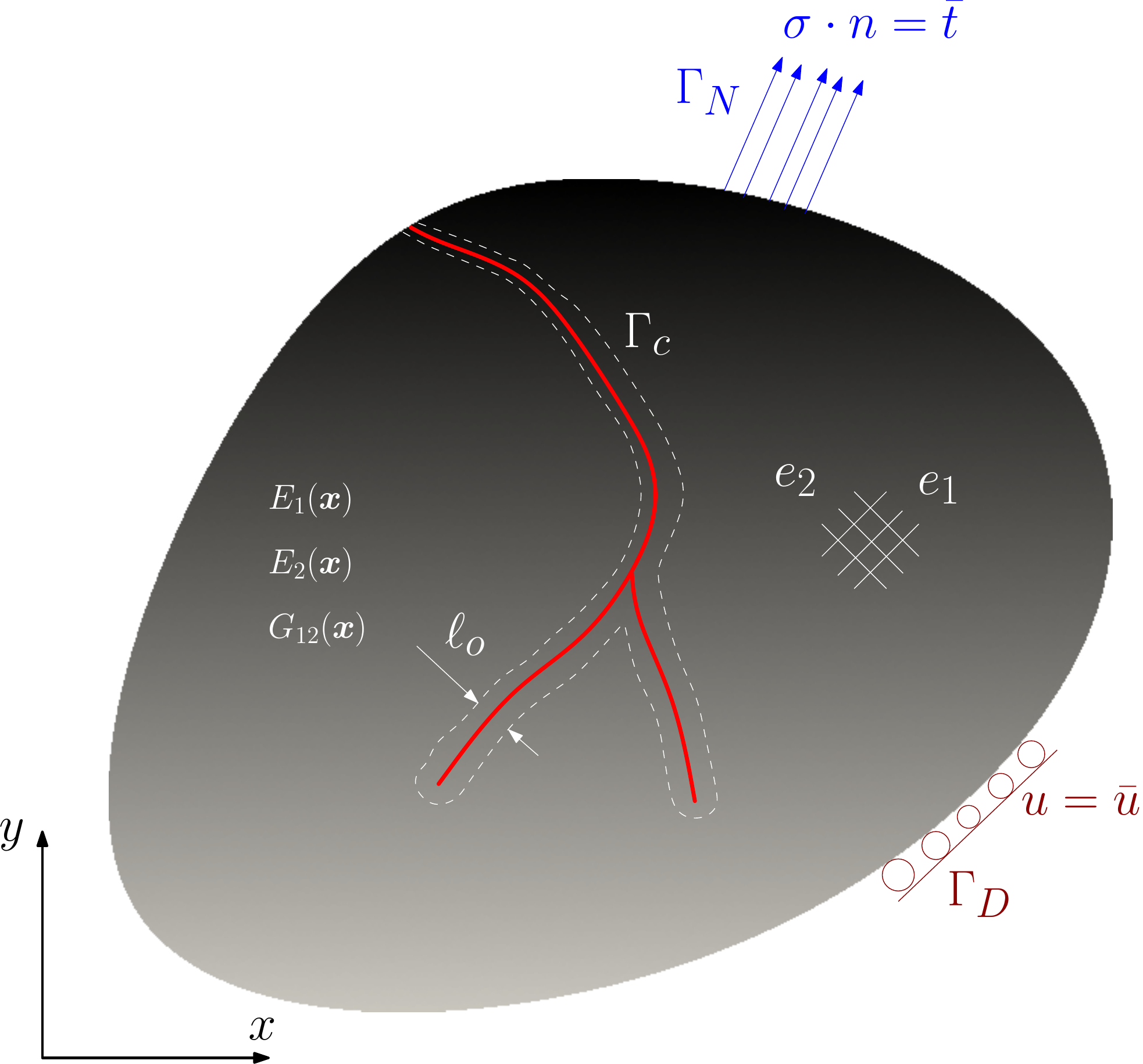}
\caption{Schematic representation of a orthotropic FGM domain with a geometric discontinuity in PFM framework, $\ell_o$ is the characteristic length scale.}
\label{domain_representation}
\end{figure}

Consider an orthotropic functionally graded solid with primary orientation directed along the axis $e_1$, making an angle $\theta$ with respect to the global frame $e_{x}$, and secondary orientation $e_2$, which is orthogonal to $e_1$ as shown in Fig. \ref{domain_representation}. The boundary ($\Gamma$) is considered to admit the decomposition with the outward normal $\cn$ into three disjoint sets, i.e., $\Gamma = \Gamma_{\rm D} \cup \Gamma_{\rm N} \cup \Gamma_{\rm c}$ and $\Gamma_{\rm D} \cap \Gamma_{\rm N} \cap \Gamma_{\rm c} = \emptyset$, where $\Gamma_{\rm c}$ is the crack surface, Dirichlet boundary and Neumann boundary conditions are specified on $\Gamma_{\rm D}$ and $\Gamma_{\rm N}$ respectively. The closure of the domain is $\overline{\Omega} \equiv \Omega \cup \Gamma$. 
\subsection{Governing equations}

The spatial variation of the elastic and fracture properties inherent to functionally graded materials (FGMs) can be incorporated following the pioneering work by Hirshikesh \textit{et al.} \cite{HIRSHIKESH2019239}. Variational phase field fracture methods are particularly suited to capture the complex crack trajectories that are observed in FGMs due to the inherent crack tip mode mixity \cite{HIRSHIKESH2019239,Doan2016,VanDo2017}. As described below, we introduce a history field $H$ to prevent damage irreversibility and we adopt the so-called hybrid model \cite{Ambati2015} to reduce the computational cost by keeping the linear form of the elasticity equation. In addition, we decompose the strain energy density into tensile and compressive parts $\psi=\psi^+ + \psi^-$, so as to prevent damage under compressive stresses. Consider a linear elastic solid with spatially varying toughness $\mathcal{G}_c(\bm{x})$ undergoing small strains. For the displacement $\uu$ and phase field $\phi$, the strong form of the governing equations in the absence of inertia and body forces is given by~\cite{Ambati2015, NGUYEN2017279}:
\begin{subequations}
 \begin{align}
 \boldsymbol{\nabla}^e\cdot\boldsymbol{\sigma}  = &~ \boldsymbol{0}   \hspace{3mm} \rm{in}  \hspace{3mm} \Omega, \label{eqn:linearmoment} \\
 -\mathcal{G}(\bm{x})_c \ell_o \nabla^p \phi \mathbb{A} \nabla^p \phi  + \left[    \frac{\mathcal{G}_c(\bm{x})}{\ell_o}  + 2 H^+ \right] \phi  = &~ 2  H^+ \hspace{3mm} \rm{in}  \hspace{3mm} \Omega, 
 \label{eqn:diffeqn}
 \end{align}
 \label{eqn:strongForm}
\end{subequations}

These balance equations are supplemented with the following boundary conditions:
 \begin{align}
 \boldsymbol{\sigma} \cdot \mathbf{n} =&~ \overline{\mathbf{t}} \quad \rm{on}  \quad \Gamma_{\rm N}, \nonumber \\
 \uu =&~ \overline{\uu} \quad \rm{on}  \quad \Gamma_{\rm D}, \nonumber \\
 \nabla \phi \cdot \mathbf{n}  =&~ 0 \quad \rm{on}  \quad \Gamma \setminus \Gamma_{\rm c},
\end{align}

\noindent where $\nabla^p$ and $\boldsymbol{\nabla}^e$ are the scalar and the vector differential operators, given by,
\begin{eqnarray}
\nabla^{p} &= \left[\begin{array}[c]{cc} \frac{\partial}{\partial x} & \frac{\partial}{\partial y}
\end{array}\right]^\mathrm {T}, \nonumber \\
\boldsymbol{\nabla}^{e} &= \left[\begin{array}[c]{ccc} \frac{\partial}{\partial x} & 0 & \frac{\partial}{\partial y} \\
    0 & \frac{\partial}{\partial y}& \frac{\partial }{\partial x} 
\end{array}\right]^ \mathrm{T},
\end{eqnarray}
Here $\mathbb{A} = \mathbf{I} + \beta\left[\mathbf{I} - \mathbf{n} \otimes  \mathbf{n}\right]$, with $\mathbf{n}=\left\{\cos\theta , \sin\theta\right\}^{\rm T}$ introduced to account for the crack path based on the material orientation. In this work, the penalty parameter, $\beta = 20$ is considered which constraint the propagation of crack in the direction perpendicular to the cleavage plane. The Cauchy stress tensor, $\bm{\sigma}$ for the functionally graded orthotropic material is defined as:
\begin{equation}
    \bm{\sigma}=\left[ (1-\phi)^2+ k_p \right] \mathbb{D}(\bm{x}) \bm{\varepsilon},
\end{equation}
where $k_p$ is a small positive number introduced for numerical stability and
\begin{equation}
    \mathbb{D}(\bm{x}) = \mathbb{T}^\mathrm{T} \bm{Q}(\bm{x}) \mathbb{T}  
\end{equation}
with 
\begin{subequations}
\begin{equation}
\mathbb{T} = \begin{bmatrix}
\cos \theta & \sin\theta & 0 \\
-\sin \theta  & \cos \theta & 0 \\ 
0 & 0 & 1
\end{bmatrix},
\end{equation} and 
\begin{equation}
\bm{Q}(\bm{x}) = \begin{bmatrix}
Q_{11} & Q_{12} & 0 \\
Q_{21} & Q_{22} & 0 \\
0 & 0 & Q_{66}
\end{bmatrix}.
\end{equation}
\end{subequations}
The components of the tensor $\bm{Q}(\bm{x})$ are calculated as:
\begin{align}
Q_{11} &= \frac{E_1(\bm{x})}{1-\nu_{12} \nu_{21}}, \qquad
Q_{22} = \frac{E_2(\bm{x})}{1-\nu_{12} \nu_{21}}, \nonumber \\
Q_{12} &= \frac{\nu_{12} E_2(\bm{x})}{1-\nu_{12} \nu_{21}} = \frac{\nu_{21} E_1(\bm{x})}{1-\nu_{12} \nu_{21}}, \nonumber \\
Q_{66} &= G_{12}(\bm{x}), \qquad
\nu_{21} = \frac{E_2(\bm{x})}{E_1(\bm{x})} \nu_{12},
\end{align}

Here, $E_1(\bm{x})$ and $E_2(\bm{x})$ are the longitudinal and the transverse Young's modulus respectively, $G_{12}(\bm{x})$ is the shear modulus, $\nu_{12}$ is the major Poisson's ratio, and $\nu_{21}$ is the minor Poisson's ratio. Thus, material properties vary at the element level, in what is usually referred to as a \emph{graded} finite element approach \cite{Materials2019}. The small strain tensor ($\bm{\varepsilon}$) is computed from the displacement field ($\vm{u}$) as, 
  \begin{equation}
      \bm{\varepsilon} = \frac{1}{2} \left( \boldsymbol{\nabla}^e \vm{u}^{\rm{T}} +  \boldsymbol{\nabla}^e \vm{u}\right).
  \end{equation}
The history variable, $H^+$ is defined as,
 \begin{equation}
     H^+:= \underset{\tau \in [0, t]}{\rm{max}} \psi^+(\bm{\varepsilon}(\bm{x}, \tau)).
     \label{Eq:H+}
\end{equation}
\noindent The introduction of $H^+$ in \eref{eqn:diffeqn} helps to decouple Equations (\ref{eqn:linearmoment}-\ref{eqn:diffeqn}) and a robust staggered scheme can be used for computing ($\bm{u}$, $\phi$)~\cite{MIEHE20102765, Ambati2015}. However, one should note that monolithic quasi-Newton methods have recently shown great promise for phase field fracture problems \cite{Wu2020a,TAFM2020}. Further, to prevent the crack faces from inter-penetration, \eref{eqn:diffeqn} is supplemented with the following constraint:
 \begin{equation}
 \forall \boldsymbol{x}: \psi^+ < \psi^- \Rightarrow \phi:= 0,
 \end{equation}
 where,
 \begin{equation*}
 \psi^{\pm}(\bm{\varepsilon}) = \frac{1}{2} \lambda \langle \rm{tr} (\bm{\varepsilon}) \rangle^{2}_{\pm}  + \mu \rm{tr} (\bm{\varepsilon}^{2}_{\pm}),
 \label{eqn:histpositivenegative}
 \end{equation*}
 with {$\langle \cdot \rangle_{\pm}:= \frac{1}{2}(\cdot \pm |\cdot|)$, $\bm{\varepsilon}_{\pm}:= \sum\limits^3_{I = 1} \langle \varepsilon_I \rangle_{\pm} \cn_I \otimes \cn_I  $ and $\bm{\varepsilon} = \sum\limits^{3}_{I = 1} \langle \varepsilon_I \rangle\cn_I \otimes \cn_I,$ where $\{\varepsilon_I \}_{I=1}^3$ and $\{\cn_I\}_{I=1}^3$ are the principal strains and the principal strain directions, respectively.}

\subsection{Weak form}
{Let $\mathscr{W}(\Omega)$ include the linear displacement field and the phase field variable, and let ($\mathscr{U}, \mathscr{P})$ and ($\mathscr{V},\mathscr{Q}$) be the trial and the test function spaces:
\begin{subequations}
 \begin{align}
 (\mathscr{U},\mathscr{V}^0) &=
 \left\{ (\uu^h,\vm{v}) \in [ C^0(\Omega)]^d : (\uu,\vm{v}) \in
 [ \mathcal{W}(\Omega)]^d \subseteq [ H^{1}(\Omega)]^d \right\},  \\
 (\mathscr{P},\mathscr{Q}^0) &= \left\{ (\phi^h,q) \in
 [ C^0(\Omega) ]^d : (\phi, q) \in [ \mathcal{W}(\Omega)]^d \subseteq
 [ H^{1}(\Omega) ]^d
 \right\}.
 \end{align}
 \end{subequations}
 Let the domain be partitioned into elements $\Omega^h$ and on using shape functions $N$ that span at least the linear space, we substitute the trial and the test functions: $\left\{\vm{u}^h, \vm{\phi}^h\right\} = \sum\limits_I N_I \left\{\vm{u}_I,\phi_I\right\}$ and $\left\{\vm{v}, q \right\} = \sum\limits_I N_I \left\{\vm{v}_I,q_I\right\}$ into \eref{equation:weakform}. The system of equations can be readily obtained upon applying the standard Bubnov-Galerkin procedure. Find $ \, \uu^h \in \mathscr{U} \, {\rm and} \, \vm{\phi}^h \in \mathscr{P} \, \text{such that, for all} \, \vm{v} \in \mathscr{V}^0 \, {\rm and} \, q \in \mathscr{Q}^0$,
  \begin{subequations}
  \begin{align}
      \int_{\Omega} \left\{ \left[ (1-\phi)^2+k_p \right]\bm{\sigma}(\vm{u}):\bm{\varepsilon}(\vm{v}) \, \right\} \text{d} 
     \Omega  =& \int_{\Gamma_t}      {\bar{t}} 
     \cdot\vm{v}\,\text{d} 
     \Gamma,
  \label{equation:elasticity}\\
      \int_{\Omega} \left[ \nabla q \, \mathcal{G}\,(\bm{x})_c \ell_o\, \mathbb{A} \nabla \phi + q \left( \frac{\mathcal{G}(\bm{x})_c}{\ell_o} + 2 H^+ \right) \phi \right] \, \text{d} \Omega =& \int_{\Omega} 2 H^+ q \, \text{d} \Omega + \int_{\Gamma} \nabla \phi \cdot \vm{n} \, q \, \text{d} \Gamma,
      \label{equation:diffusion}
  \end{align}
  \label{equation:weakform}
  \end{subequations}
 }
\noindent which leads to the following system of linear equations:
 \begin{subequations}
     \begin{align}
     \mathbf{K}^{\rm uu} \vm{u}^h &= \vm{f}^{\rm uu}, \label{eqn:elast_discrete} \\
     \mathbf{K}^{\phi} \vm{\phi}^h &= \vm{f}^{\phi}, \label{eqn:phase_discrete}
 \end{align}
 \end{subequations}
 where
  \begin{align*}
      \mathbf{K}^{\rm uu} &= \sum\limits_h \int\limits_{\Omega^h} \Big[(1-\phi)^2+k_p\Big] \mathbf{B}^{\rm T}~\mathbb{D}(\bm{x})~\mathbf{B}~\mathrm{d}\Omega, \\
      \mathbf{K}^{\phi} &= \sum\limits_h \int\limits_{\Omega^h} \Bigg[ \mathbf{B}^{\rm T}_{\phi}~\mathcal{G}(\bm{x})_c \ell_o \mathbb{A}~\mathbf{B}_{\phi} + \mathbf{N}^{\rm T}\left( \frac{\mathcal{G}(\bm{x})_c}{\ell_o} + 2 H^+ \right) \mathbf{N} \Bigg]~\mathrm{d}\Omega, \\
      \vm{f}^{\rm uu} &= \sum\limits_h \int\limits_{\Omega^h}  \mathbf{N}^{\rm T} {\bar{t}}~\mathrm{d}\Omega, \\
      \vm{f}^{\phi} &= \sum\limits_h \int\limits_{\Omega^h} \mathbf{N}^{\rm T}~2 H^+~\mathrm{d}\Omega,
  \end{align*}
Here, $\mathbf{B}= \boldsymbol{\nabla}^e \mathbf{N}$ is the strain-displacement matrix and $\mathbf{B}_{\phi} = \nabla^p \mathbf{N}$ is the scalar gradient of the shape function matrix $\mathbf{N}$. The above system of equations are solved by the staggered approach \cite{MIEHE20102765, Ambati2015}. The present framework is implemented in Matlab. The reader is referred to Ref. \cite{Hirshikesh2019380} for a FEniCS-based implementation, Ref. \cite{ZHOU201831} for a COMSOL-based implementation, Ref. \cite{Msekh2015} for an Abaqus-based implementation, and to Ref. \cite{Samaniego2020112790} for machine learning solution scheme.

\section{Recovery based error indicator and quadtree decomposition} \label{Sec:Quadtree Meshing}
In this section, we present a brief overview of the recovery based error indicator proposed by Bordas and Duflot \cite{Bordas2007,Bordas_errror_ind_XFEFM2008} for the XFEM. This is done to assess the error and identify the elements/regions which have to be refined. Later, the process of quadtree decomposition is discussed.

\subsection{Recovery based error indicator}
\begin{figure}[!htbp]
	\centering 
 \includegraphics[width=0.7\textwidth]{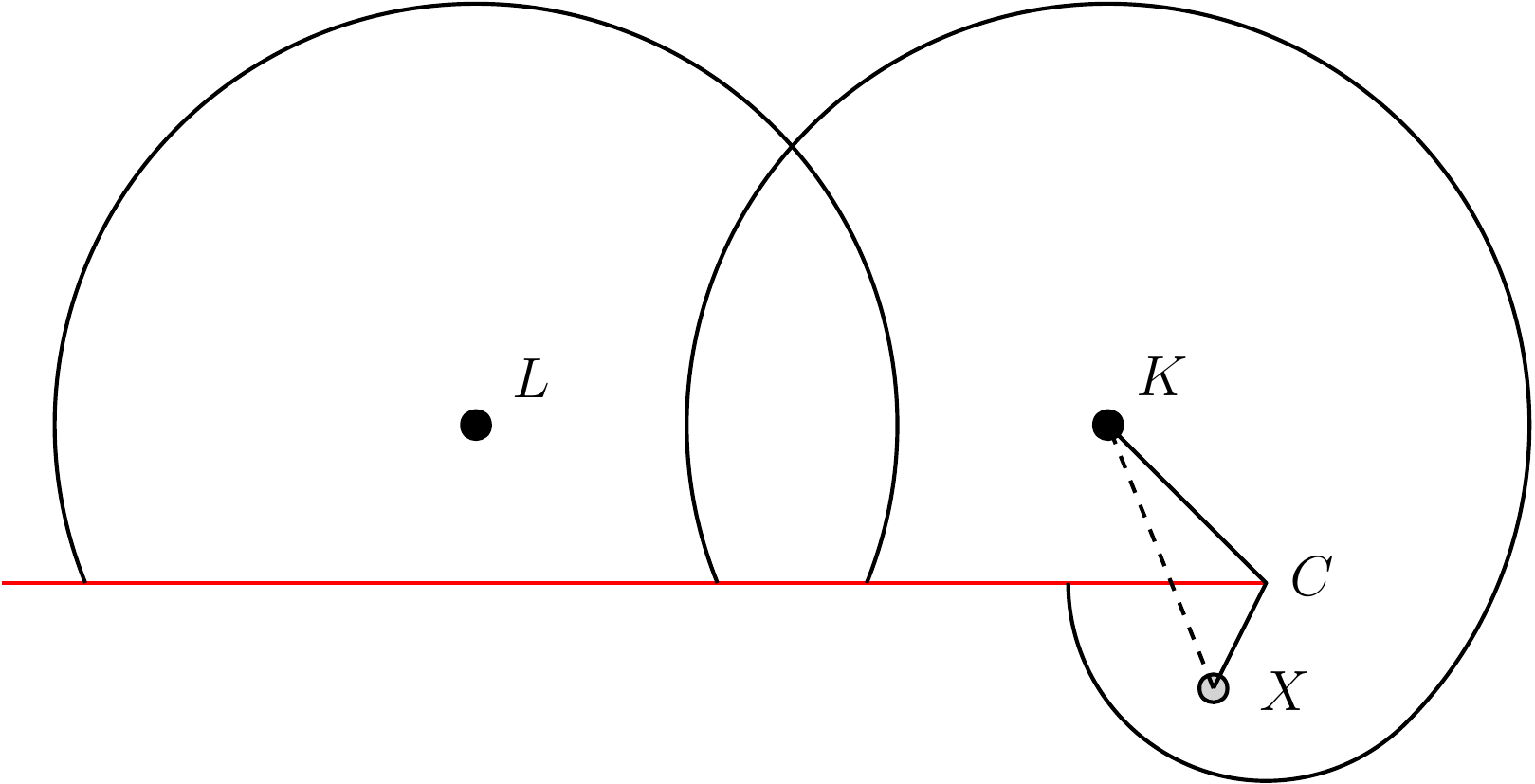}
\caption{Diffraction method for calculating weight function considering discontinuities in moving least-squares approximations (red line shows the discontinuity).}
\label{fig:diffraction method}
\end{figure}

In this method, the enhanced strain field is computed using the standard nodal solution through the eXtended Moving Least Square (XMLS) derivative recovery process. This is then further used as error indicator. Let $\mathbf{x}$ be a point in the domain, and $n_{x}$ XMLS points contain $\mathbf{x}$ in their domain of influence. Then, using the displacement values at these $n_{x}$ points, the enhanced displacement field and the strain field at $\mathbf{x}$ can be written as,
\begin{equation}
\label{eqn:enhanced displacement}
{\vm{u}^s(\mathbf{x})=\sum_{I=1}^{n_{x}} {{\psi}_{k}(\mathbf{x})}{\vm{u}^h_k} = {\mathbf{\Psi}^{\rm{T}}(\mathbf{x})}{\vm{u}^h},} 
\end{equation}
\begin{equation}
\label{eqn:enhanced strain}
{\bm{\varepsilon}^s(\mathbf{x})=\sum_{I=1}^{n_{x}} \mathscr{D}({{\psi}_{k})(\mathbf{x})}{\vm{u}^h_k} = {{\mathbf{D}}(\mathbf{x})}{\vm{u}^h},}  
\end{equation}
where ${\Psi}_{k}(\mathbf{x})$ is the MLS shape function value associated with node k at $\mathbf{x}$, $\mathscr{D}$ is the derivative operator and $\mathbf{D}$ is the MLS shape function derivative matrix. The matrix form of the MLS shape function is given by:
\begin{equation}
\label{eqn:MLS shapefunction}
\mathbf{\Psi}^{\rm T}(\mathbf{x})=
\begin{bmatrix}
\psi_{1}(\mathbf{x}) &
\psi_{2}(\mathbf{x}) & ... &  \psi_{n_{x}}(\mathbf{x})
\end{bmatrix}
=\mathbf{p}^{\rm T}(\mathbf{x}) \mathbf{A}^{-1}(\mathbf{x}) \mathbf{B}(\mathbf{x}), 
\end{equation}
where $\mathbf{p}(\mathbf{x})$ denotes the $m$ reproducing polynomial used for the MLS shape function. For two dimensions, $\mathbf{p}(\mathbf{x}) = [ 1 \quad x \quad y ]$, and,
\begin{align*}
\mathbf{A}(\mathbf{x}) &=\sum_{I=1}^{n_{x}} w_I(\mathbf{x}) \mathbf{p}(\mathbf{x}_{I}) \mathbf{p}^{\rm T}(\mathbf{x}_{I}) \nonumber \\
\mathbf{B}(\mathbf{x}) &=
\begin{bmatrix}
w_1(\mathbf{x}) \mathbf{p}(\mathbf{x}_{1}) &
w_2(\mathbf{x}) \mathbf{p}(\mathbf{x}_{2}) & ... &  w_{n_{x}}(\mathbf{x}) \mathbf{p}(\mathbf{x}_{n_{x}})
\end{bmatrix}.
\end{align*}
Here, $\mathbf{A}$ is a $m \times m$ matrix and $\mathbf{B}$ is a $m \times n_x$ matrix. For the matrix $\mathbf{A}$ to be invertible, we need $n_x > m$, i.e., we need more number of points whose domains of influence contains $x$ that the basis functions in $\mathbf{p}$. However, note that this is not a sufficient condition. The weight function $\mathbf{w}_I$ associated with a node $\mathbf{x}_I$ is calculated by the diffraction method with a circular domain of influence. The domain of influence also changes if it intersects with the discontinuity. In this work, a fourth order spline is taken as the weighting function \cite{Bordas_errror_ind_XFEFM2008}:
\begin{equation}
w_k(\mathbf{x}) = 
\begin{cases}
    1- 6{s^2}+8{s^3}-3{s^4} & \text{if} \hspace{0.3cm} \vert s \vert \leq 1  \\ 
	 0\hspace{1cm}               & \text{if} \hspace{0.3cm}\vert s \vert > 1
\end{cases}.
\end{equation}
where $s=\frac{\Vert \mathbf{x}-\mathbf{x}_{k}\Vert}{d_{k}}$ and $d_k$ denotes the support domain of node $\mathbf{x}_k$. $s$ is calculated differently than above to account for a discontinuity in the approximation. When describing a discontinuity, if it covers a point, a node's weight at this point will decrease. When the line section $C_{i}X$ of \fref{fig:diffraction method} is bisected by a crack, $s$ of \eref{eqn:diffraction_for_discont} is substituted by the normalized length of the shortest path from K to X that passes through a front point (route KCX in \fref{fig:diffraction method}).
\begin{equation}
s=\frac{\Vert \mathbf{x}-\mathbf{x}_{c}\Vert + \Vert \mathbf{x}_{c}-\mathbf{x}_k\Vert}{d_{k}}
\label{eqn:diffraction_for_discont}
\end{equation}

The enhanced derivatives of the shape functions are computed by finding the derivatives of the MLS shape functions, see \eref{eqn:MLS shapefunction}. The enhanced derivatives of the displacements and the enhanced small strain $\bm{\varepsilon}_s$ can then be found. The error is computed by comparing the enhanced strain field to the standard compatible strain field. Thus, the total error of domain $\Omega$ is,
\begin{equation}
   \Vert \mathbf{e} \Vert_{\Omega} = \sqrt{\int_{\Omega} \Vert \bm{\varepsilon}(\mathbf{x}) -  \bm{\varepsilon}_s(\mathbf{x}) \Vert^{2}~  \text{d} \mathbf{x}} 
\end{equation}
while the error within the element $i$ with area $\Omega_{i}$ is,  
\begin{equation}
   \Vert \mathbf{e} \Vert_{\Omega_{i}} = \sqrt{\int_{\Omega{i}} \Vert \bm{\varepsilon}(\mathbf{x}) -  \bm{\varepsilon}_s(\mathbf{x}) \Vert^{2}~  \text{d} \mathbf{x}} 
\end{equation}


The tolerance is chosen based on the maximum error criteria. Thus, the elements with high individual error are discretized in the next level; all elements whose individual error is higher than given tolerance value will be sent for discretization in the next level.

\subsection{Quadtree decomposition}
\begin{figure}[H]
    \centering
 \begin{subfigure}[h]{0.3\textwidth}
                \centering                \includegraphics[width=0.9\textwidth]{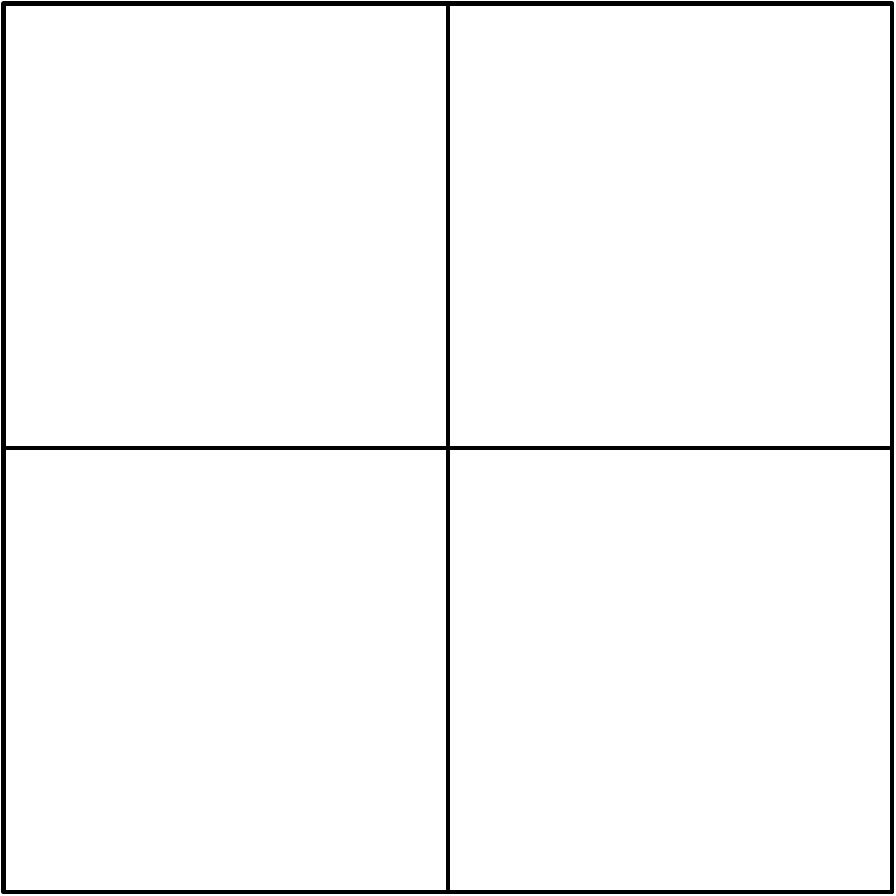}
                \caption{}
                \label{fig:Meanvalueshapefunctiona}
\end{subfigure}
\begin{subfigure}[h]{0.3\textwidth}
               \centering                \includegraphics[width=0.9\textwidth]{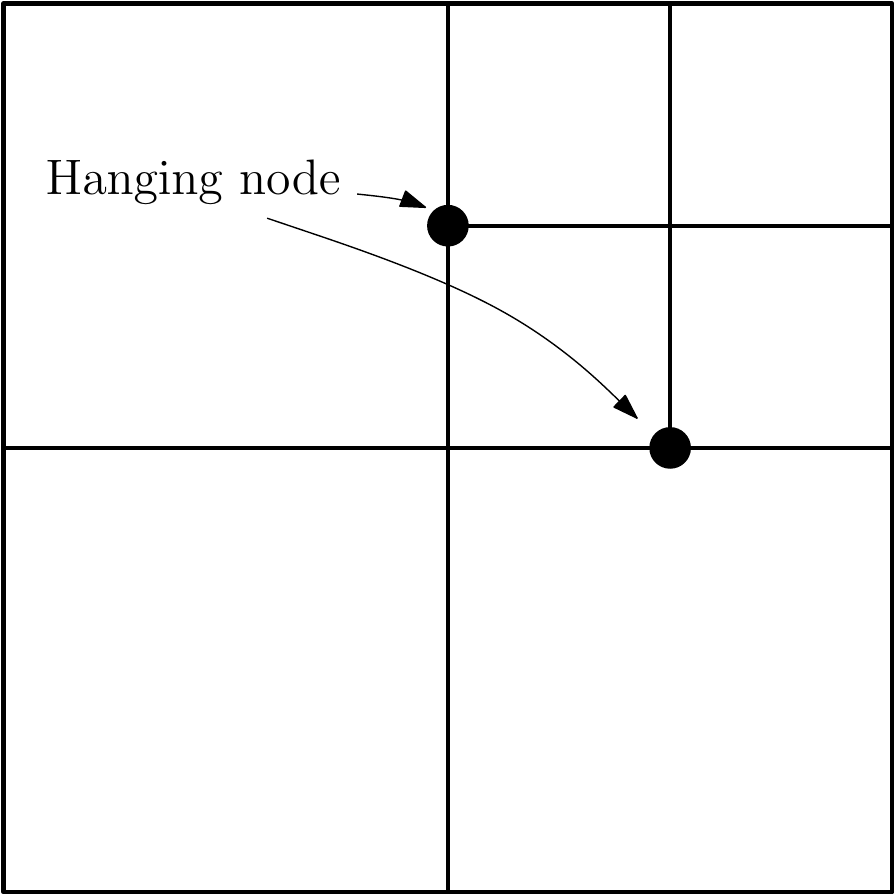}
                \caption{}
               
\end{subfigure}
\begin{subfigure}[h]{0.3\textwidth}
               \centering                \includegraphics[width=0.9\textwidth]{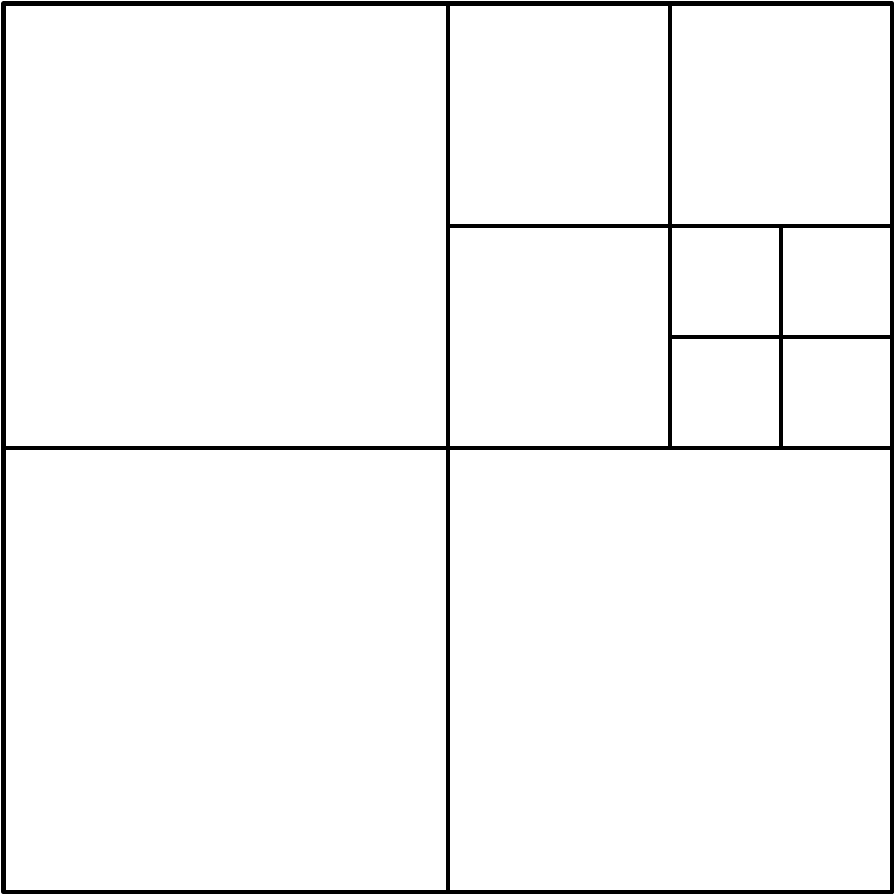}
                \caption{}
               
\end{subfigure}
\caption{Quadtree decomposition: (a) representative quadtree mesh and (b) tree structure employed to store the mesh details.}
\label{fig:Quadtree mesh details}
\end{figure}

The quadtree decomposition is used for local refinement once the error is quantified. The quadtree decomposition entails several features; namely, (a) is easy to implement (b) it requires less degrees of freedom (Dofs) and (c) retains hierarchical mesh structures. The hierarchical mesh structure facilitates efficient computations, particularly efficient storage and data retrieval. In this decomposition, the so-called stopping criterion is used to decide which element requires to be further refined. This criterion can be a geometry based factor or any error indicator. The criteria for an element to be refined based on the error indicator could be based on either equal distribution criterion or Min-number criteria~\cite{JIN2017319}. In this work, equal distribution criterion is used to minimize the global error and balance the local error throughout the domain. If the given element does not satisfy the stopping criterion within the user specified tolerance limit, it will be divided into four child elements as shown in \fref{fig:Quadtree mesh details}. This process can be repeated several times until the criteria is met. The tolerance in all the examples is chosen to be $1 \times 10^{-5}$.

The aforementioned decomposition leads to elements with hanging nodes; see, \fref{fig:Quadtree mesh details}. The conventional finite element approach cannot handle such elements without additional work. This is because of lack of compatibility between the elements. To restrict the number of hanging nodes per edge, a general practice 2:1 rule is applied, in which the mesh is constructed in such a way that two neighboring elements do not differ by more than one level. A number of techniques to handle these hanging nodes have been proposed, such as triangulation \cite{Greaves1999}, transformation of the hanging degrees of freedom to corner degrees of freedom using constraint equations~\cite{Fries2011}, use of special conforming shape functions~\cite{Gupta1978}, considering the element with hanging nodes as a polygon~\cite{TABARRAEI2005686}, or the use of other advanced methods like Scaled Boundary Finite Element Method (SBFEM) or Smoothed Finite Element Method (SFEM)~\cite{Natarajan2015}.

In this work, the elements with hanging nodes are considered as  $n$-sided polygons (see~ Fig. \ref{fig:Meanvalueshapefunctiona}). The mean-value shape functions proposed by Floater~\cite{Floater2003_meanvalue} are used to approximate the unknown fields. The reasons behind this choice are the facts that some angles of elements with hanging nodes are $180^\circ$ and the mean value shape-functions work efficiently for non-convex polygons.

\begin{figure}[H]
    \centering
 \begin{subfigure}[h]{0.45\textwidth}
                \centering                \includegraphics[scale =0.4]
 {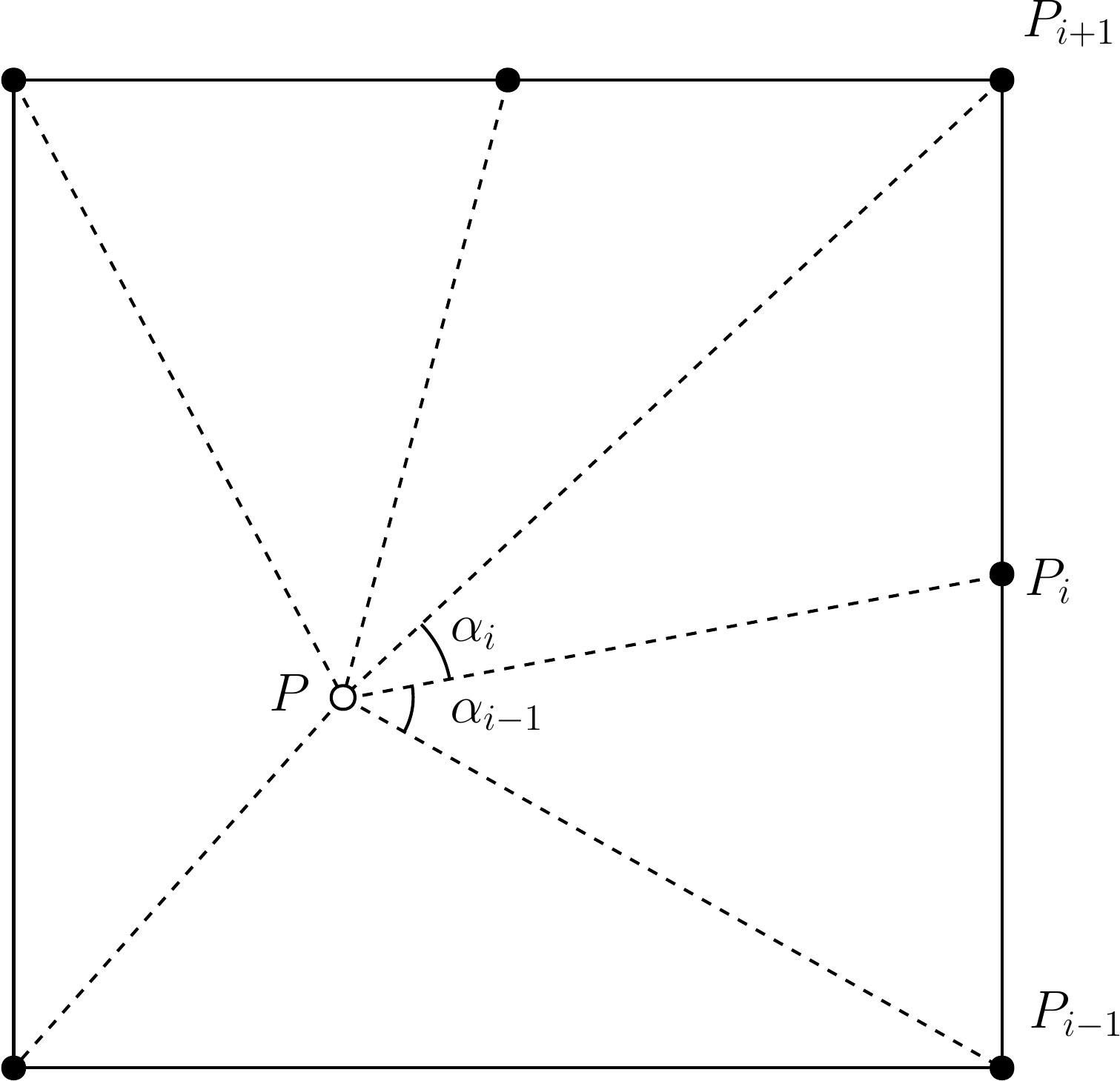}
                \caption{}
                \label{fig:Meanvalueshapefunctiona}
\end{subfigure}
\begin{subfigure}[h]{0.45\textwidth}
               \centering       
               \includegraphics[scale = 0.5]{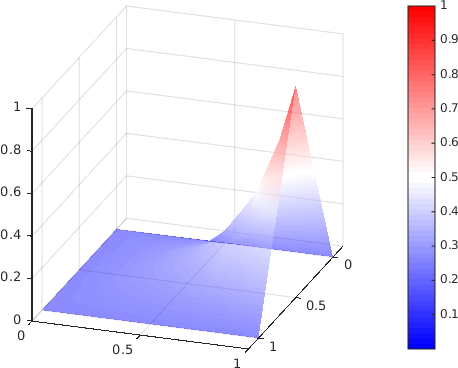}
                \caption{}
               \label{fig:Meanvalueshapefunctionb}
                \end{subfigure}
\caption{Schematic representation of an element with hanging node and the construction of mean value shape function.}
\label{fig:Meanvalueshapefunction}
\end{figure}

The mean-value coordinates for a point $P(\mathbf{x})$ in an arbitrary polygon are given by:
\begin{align}
N_{i}(\mathbf{x})=&\dfrac{\omega_{i}(\mathbf{x})}{\sum_{j=1}^{n} \omega_{i}(\mathbf{x})}, \quad i=1,\cdots,n \nonumber \\
\omega_{i}(\mathbf{x})=& \dfrac{\tan(\alpha_{i-1}/2) + \tan(\alpha_{i}/2)}{\Vert \mathbf{x}-\mathbf{x}_{i}\Vert} 
\label{meanvalueshapefunction}
\end{align}
where $n$ is the number of nodes in an element, $\mathbf{x}_{i}$ are the coordinates of point ${P}_{i}$ and $\alpha_i$'s are the internal angles. \fref{fig:Meanvalueshapefunctionb} shows the mean value shape function for the polygon with hanging node. The numerical integration for the polygonal elements is performed by subdividing the polygon into triangles and employing standard quadrature rule.

\section{Results}
\label{Sec:Results}
In this section, the performance and the robustness of the adaptive PFM for fracture of orthotropic FGMs is investigated. We first validate the adaptive PFM results against experimental and numerical results for failure of orthotropic materials. Then, cracking of orthotropic FGMs is investigated for different material grading possibilities. The numerical stability parameter $k_p$ is assumed to be 1 $\times 10^{-6}$ in all the numerical examples, unless specified otherwise. The proposed adaptive PFM is implemented in MATLAB R2014b and the simulations were performed on Intel quad Core i5-4590CPU@3.30 GHz with 8 GiB RAM. 

\subsection{Validation: fracture of orthotropic materials}

\label{section:result_orthotropic}
\begin{figure}[H]
    \centering
 \begin{subfigure}[h]{0.4\textwidth}
                \centering                \includegraphics[scale=0.5]{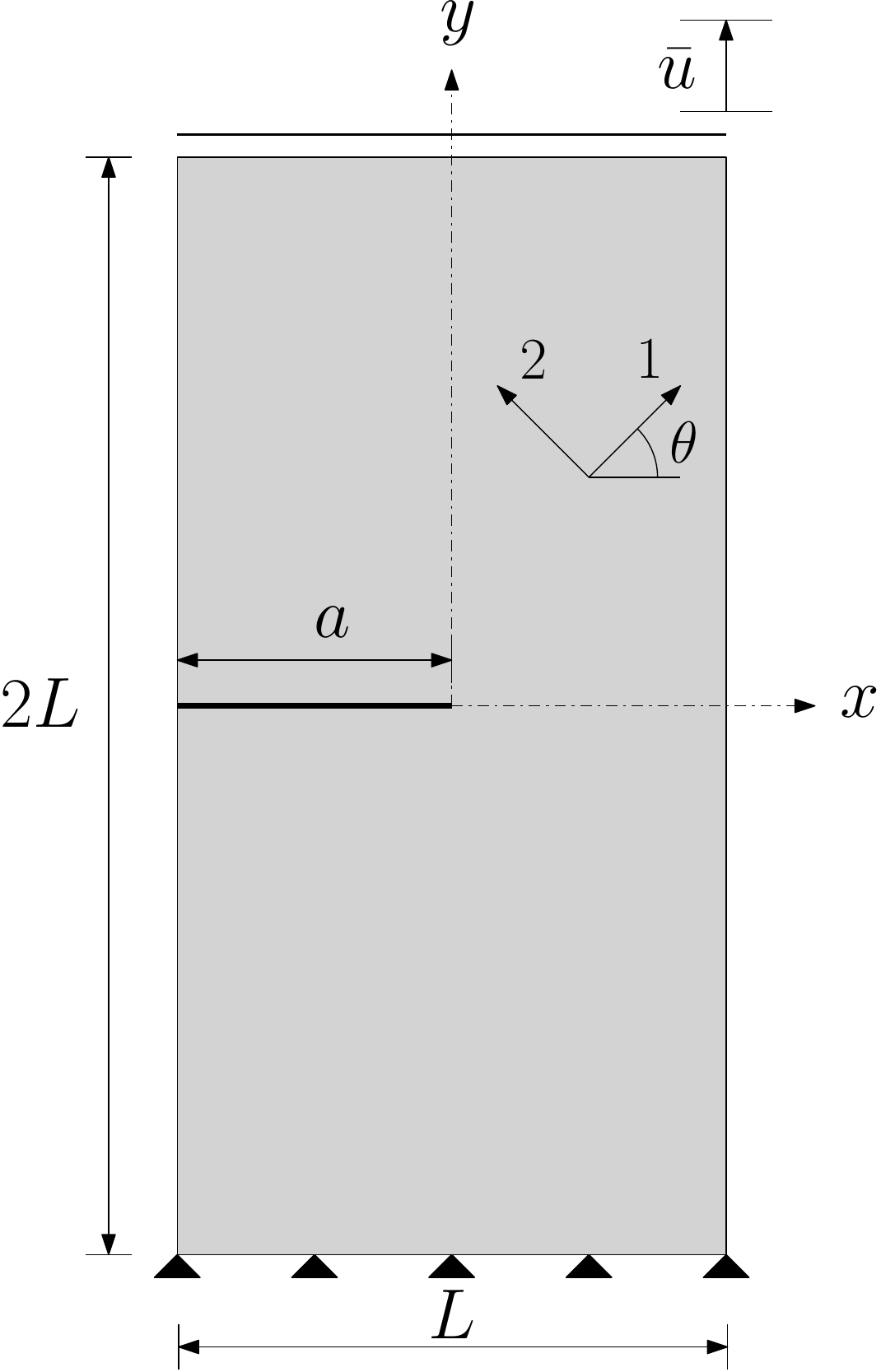}
                \caption{}
                \label{fig:orthortropic_domian}
\end{subfigure}
\begin{subfigure}[h]{0.4\textwidth}
               \centering       
               \vspace{2mm} \includegraphics[scale=0.25]{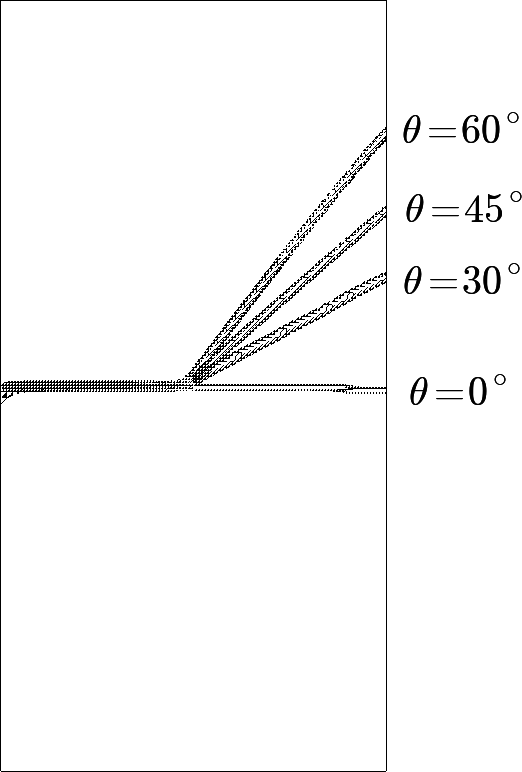}
                \caption{}
                \label{fig:orthortropic_crackpropagation}
                \end{subfigure}
\caption{Edge crack orthotropic specimen: (a) geometry, material properties and boundary conditions (b) crack propagation direction for different material orientation. [where $L =$ 1 mm and $a = $ 0.5 mm].}
\end{figure}

The framework developed is validated first with the experimental and numerical work (XFEM) of Cahill {\it et al., }\cite{Cahill2013119}. In order to study the fracture processes in an orthotropic material, an edge crack specimen subjected to tensile loading is considered, see Figure \ref{fig:orthortropic_domian}. The material properties are chosen as: $E_1 = 114.8$ GPa, $E_2=11.7$ GPa, $G_{12} = 9.66$ GPa, $\nu_{12} = 0.21$ and the critical toughness $\mathcal{G}_c = 2.7$ MPa mm.

\begin{figure}[H]
        \centering
        \begin{subfigure}[h]{0.2\textwidth}
                \centering
                \includegraphics[scale=1]{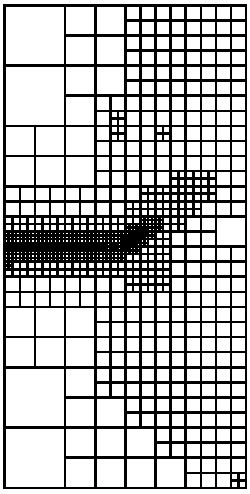}
                \caption{}
                \label{fig:crackpath1}
        \end{subfigure}
        \begin{subfigure}[h]{0.2\textwidth}
                \centering
                \includegraphics[scale=1]{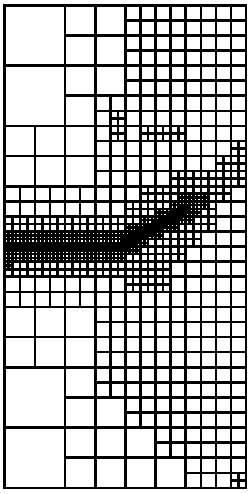}
                \caption{}
                \label{fig:crackpath2}
        \end{subfigure}
        \begin{subfigure}[h]{0.2\textwidth}
                \centering
                \includegraphics[scale=1]{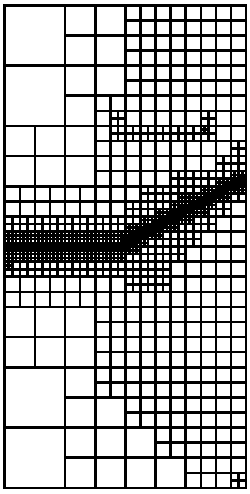}
                \caption{}
                \label{fig:crackpath2}
        \end{subfigure}
        
        \vspace{7mm}
        \begin{subfigure}[h]{0.2\textwidth}
                \centering
                \includegraphics[scale=0.2]{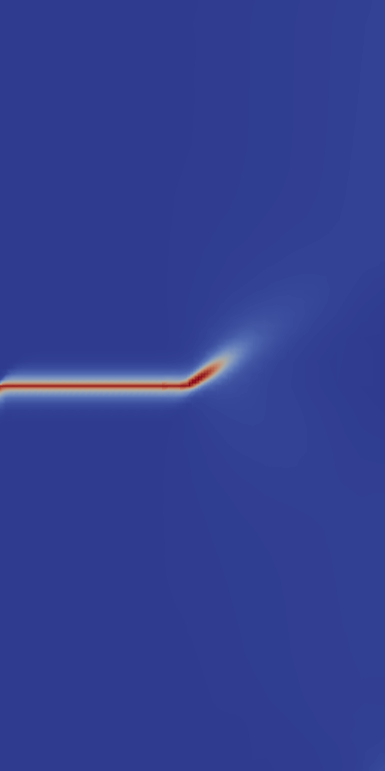}
                \caption{}
                \label{fig:crackpath1}
        \end{subfigure}
        \begin{subfigure}[h]{0.2\textwidth}
                \centering
                \includegraphics[scale=0.2]{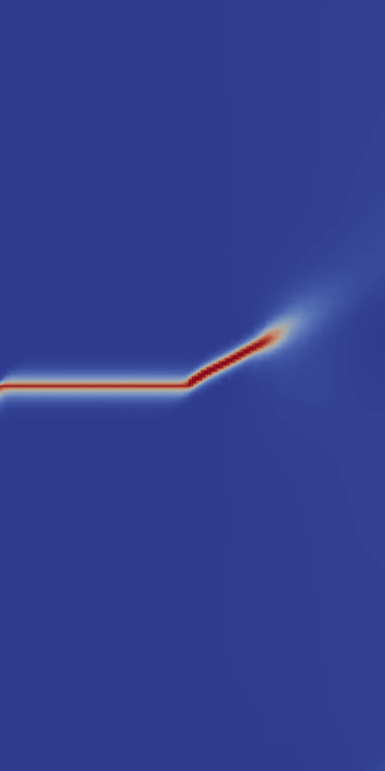}
                \caption{}
                \label{fig:crackpath2}
        \end{subfigure}
        \begin{subfigure}[h]{0.2\textwidth}
                \centering
                \includegraphics[scale=0.2]{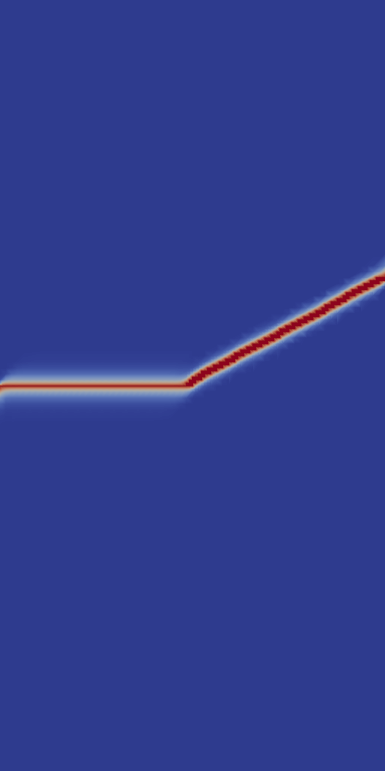}
                \caption{}
                \label{fig:crackpath2}
        \end{subfigure}
        \begin{subfigure}[h]{\textwidth}
                \centering
                \includegraphics[scale=0.45]{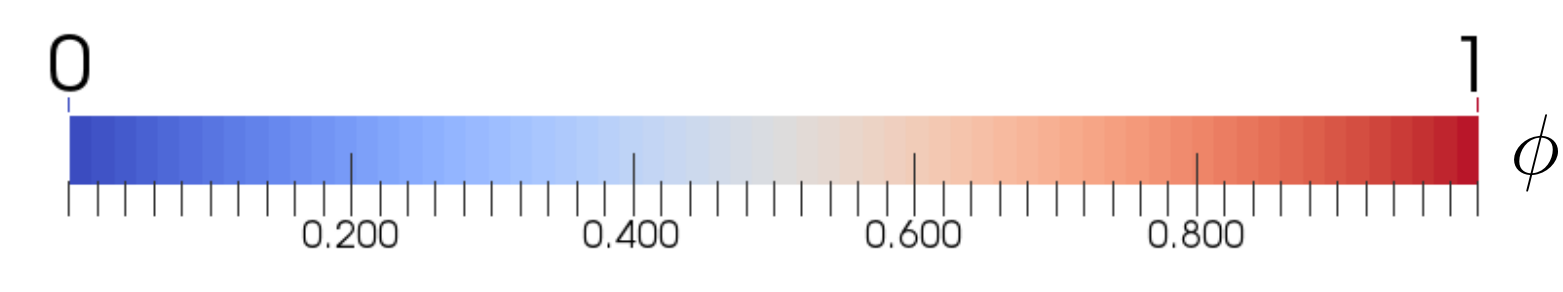}
        \end{subfigure}
        \caption{Domain discretization of edge crack specimen with $\theta = 30^\circ$ at (a) 0.035 (b) 0.038 and (c) 0.039 mm.}\label{fig:domain_discretization_30}
\end{figure}
The simulation starts with a coarse mesh and an assumed characteristic length scale, $\ell_o$. For each load step, the domain is discretized as explained in Section \ref{Sec:Quadtree Meshing}, which allows to track the crack trajectory continuously. Figure \ref{fig:domain_discretization_30} shows the domain discretization for the evolving crack trajectory; the combination of quadtree decomposition and post-errori error estimator strategies leads to a fine discretization in the vicinity of the propagating crack tip. This feature substantially reduces the size of the global stiffness matrix, resulting in reduced CPU memory requirement and efficient computations.

We proceed to quantify the computational gains by comparing to the case of uniform refinement. Computation times are compared to those obtained from a solution with uniform mesh, where the characteristic element size equals the element size in the crack region in the adaptive re-meshing case. Results are shown in \tref{table:timecompare} for the initial load step $\Delta u$, after convergence has been achieved. The number of degrees of freedom (DOFs) is also shown. Very substantial computationally gains are reported relative to the uniform mesh scenario, due to the significantly smaller number of DOFs required when using an adaptive mesh refinement strategy. The error indicator is the most time consuming operation in the adaptive PFM.

\begin{table}[!hbtp]
\caption{Computational time (in seconds) comparison for adaptive PFM and PFM with uniform refinement.}
\label{table:timecompare}
\begin{adjustbox}{max width=\textwidth}
\begin{tabular}{cccccccc}
\hline
\multirow{2}{*}{PFM strategy} & \multirow{2}{*}{DOFs} & \multicolumn{6}{c}{computation times (sec)}                                                             \\ \cline{3-8} 
                              &                       & Error indicator & remeshing & assemble ($\phi$) & soln ($\phi$) & assembly ($\bm{u}$) & soln ($\bm{u}$) \\ \hline
Adaptive refinement           & 2,772                 & 4.2             & 0.74      & 0.48              & 0.004         & 0.62                & 0.012           \\
Uniform refinement            & 108,336               & -               & -         & 11.3              & 0.30          & 17.04               & 1.48            \\ \hline
\end{tabular}
\end{adjustbox}
\end{table}

\begin{figure}[H]
        \centering
        \begin{subfigure}[h]{0.2\textwidth}
                \centering
                \includegraphics[scale=1]{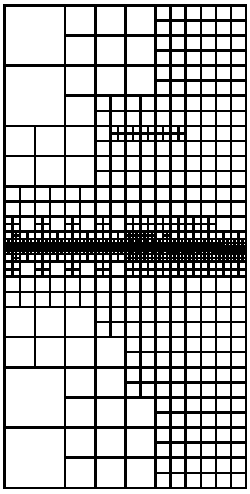}
                \caption{}
                \label{fig:crackpath1}
        \end{subfigure}
        \begin{subfigure}[h]{0.2\textwidth}
                \centering
                \includegraphics[scale=1]{th30.pdf}
                \caption{}
                \label{fig:crackpath2}
        \end{subfigure}
        \begin{subfigure}[h]{0.2\textwidth}
                \centering
                \includegraphics[scale=1]{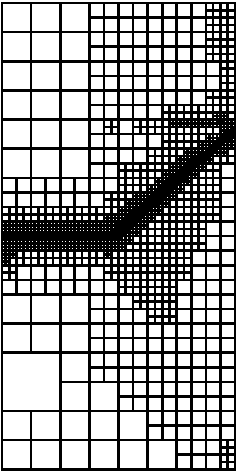}
                \caption{}
                \label{fig:crackpath3}
        \end{subfigure}
        \begin{subfigure}[h]{0.2\textwidth}
                \centering
                \includegraphics[scale=1]{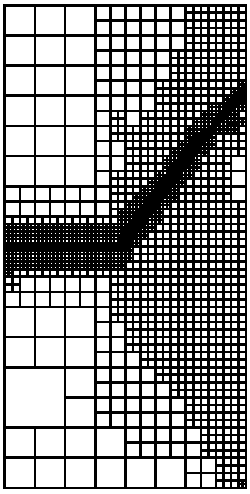}
                \caption{}
                \label{fig:crackpath3}
        \end{subfigure}
        
        \vspace{5mm}
        \begin{subfigure}[h]{0.2\textwidth}
                \centering
                \includegraphics[scale=0.2]{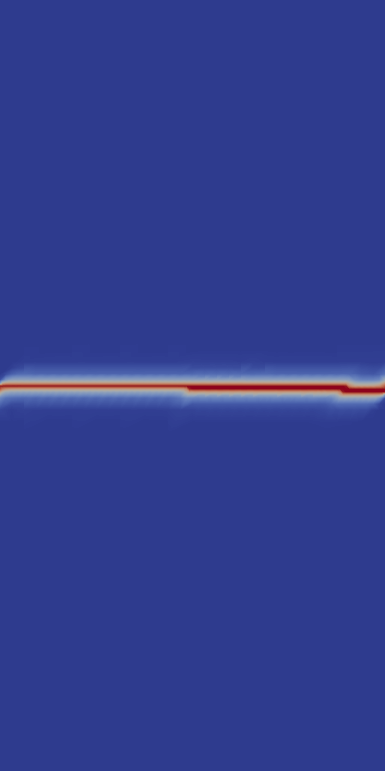}
                \caption{}
                \label{fig:crackpath1}
        \end{subfigure}
        \begin{subfigure}[h]{0.2\textwidth}
                \centering
                \includegraphics[scale=0.2]{Ortho30200.png}
                \caption{}
                \label{fig:crackpath2}
        \end{subfigure}
        \begin{subfigure}[h]{0.2\textwidth}
                \centering
                \includegraphics[scale=0.2]{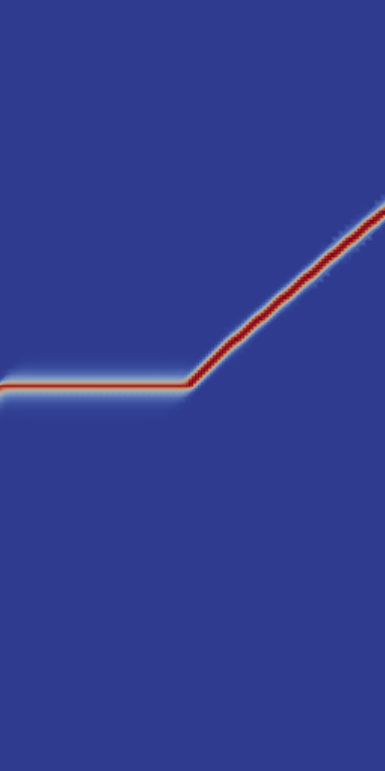}
                \caption{}
                \label{fig:crackpath3}
        \end{subfigure}
        \begin{subfigure}[h]{0.2\textwidth}
                \centering
                \includegraphics[scale=0.2]{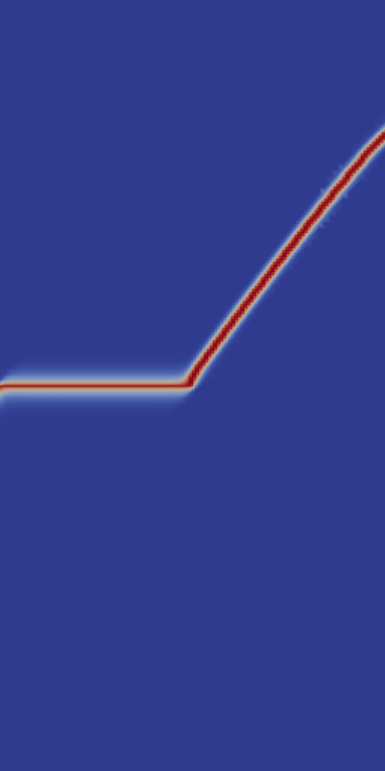}
                \caption{}
                \label{fig:crackpath3}
        \end{subfigure}
        \begin{subfigure}[h]{\textwidth}
                \centering
                \includegraphics[scale=0.45]{cbar-eps-converted-to.pdf}
        \end{subfigure}
        \caption{Final domain discretization for (a) $\theta = 0^\circ$ (b) $\theta = 30^\circ$ (c) $\theta = 45^\circ$ and (d) $\theta = 60^\circ$ and the corresponding crack trajectory in (e,f,g,h), respectively.  }\label{fig:cracking_all_angle}
\end{figure}

Figure \ref{fig:orthortropic_crackpropagation} shows the crack propagation trajectory for four selected values of material orientation i.e., $\theta = 0^\circ, 30^\circ, 45^\circ, 60^\circ$. The corresponding domain discretization is shown in Fig. \ref{fig:cracking_all_angle}. In agreement with expectations, the crack propagation path strictly follows the material orientations, see Fig. \ref{fig:orthortropic_crackpropagation}. 

\begin{figure}[H]
    \centering
    \includegraphics[scale=1]{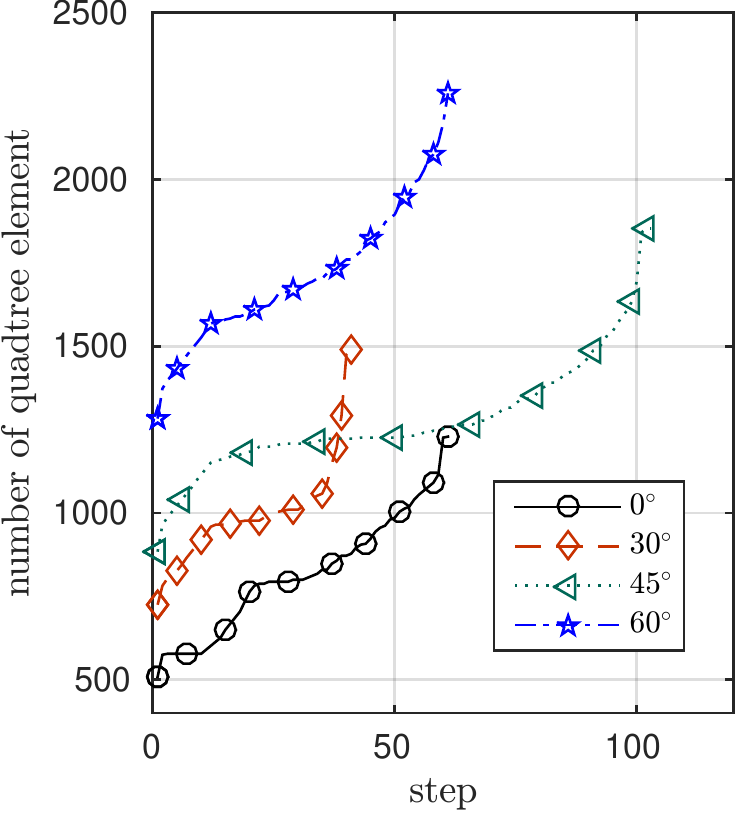}
    \caption{Quadtree evolution as a function of load step.}
    \label{fig:dof_evolution}
\end{figure}

Fig. \ref{fig:dof_evolution} shows how the number of quadtree elements increases as the crack propagates. It is shown that the material orientation angles that translate into a larger crack deflection require a larger number of quadtree elements. The comparison with the results obtained in experiments and XFEM-based calculations \cite{Cahill2013119} are shown in Table \ref{compare_Angle}. A very good agreement is attained for all values of material orientation.

\begin{table}[H]
\centering
\caption{Crack propagation angle compared with the experiments and the XFEM}
\label{compare_Angle}
\begin{tabular}{lllll}
\hline
\multicolumn{1}{l}{Fiber Orientation (degree)} & \multicolumn{1}{l}{0$^\circ$} & \multicolumn{1}{l}{30$^\circ$} & \multicolumn{1}{l}{45$^\circ$} & \multicolumn{1}{l}{60$^\circ$} \\ \hline
$\theta_{\rm{inc}}$ Experimental~\cite{Cahill2013119}        & 0                      & 30                     & 45                      & 60                      \\
$\theta_{\rm{inc}}$ XFEM~\cite{Cahill2013119}                & 0                      & 29                      & 43                      & 57                      \\
$\theta_{\rm{inc}}$ present PFM         & 0                      & 29.1                      & 43                      & 55                     \\ \hline
\end{tabular}
\end{table}

\begin{figure}[H]
    \centering
    \includegraphics[scale=1.2]{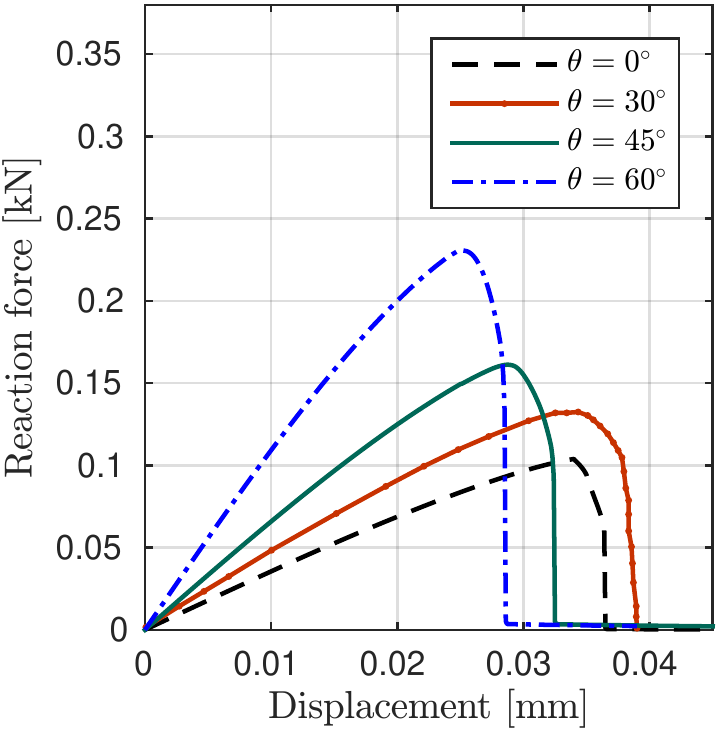}
    \caption{Load displacement response for the orthotropic material with different material orientation, $\theta$.}
    \label{fig:load_disp_edge}
\end{figure}

Finally, Figure \ref{fig:load_disp_edge} shows the load-displacement response of the orthotropic specimen with different material orientations. The stiffness of the response increases with the material orientation angle $\theta$.

\subsection{Fracture of orthotropic functionally graded materials}
\begin{figure}[H]
    \centering
    \includegraphics[scale=0.5]{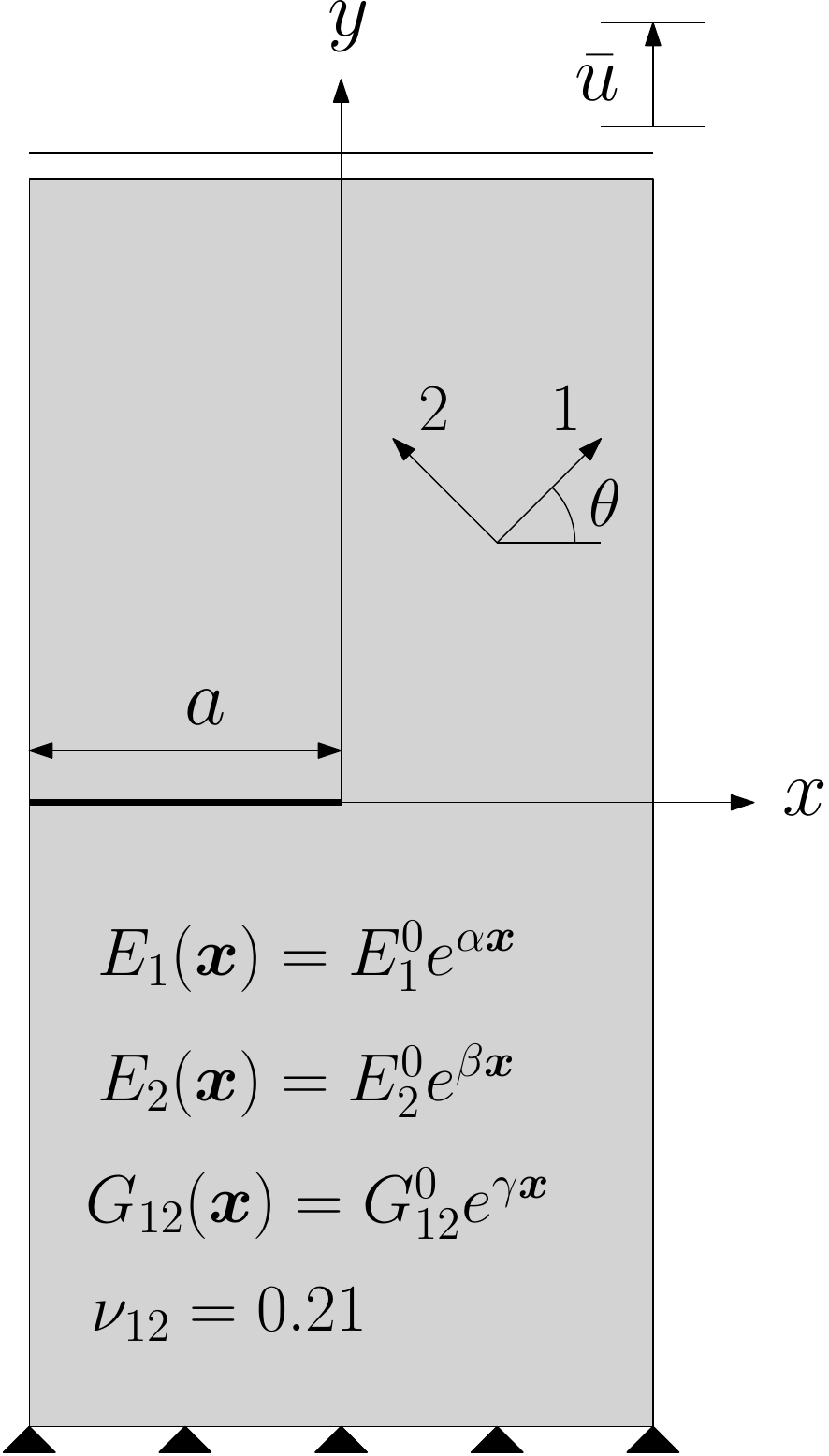}
    \caption{Orthotropic FGM specimen: domain and boundary conditions, $(\alpha, \beta, \gamma)$ are indices that control the material variation.}
    \label{fig:domian_edge_ortho}
\end{figure}

Next, we examine the fracture processes in an orthotropic FGM specimen as shown in Fig. \ref{fig:domian_edge_ortho}. In terms of material gradation, we consider the following representative case studies:
\begin{itemize}
    \item plate with a crack parallel to the material gradation i.e., $x-$direction grading,
    \item plate with a crack perpendicular to the material gradation i.e., $y-$direction grading.
\end{itemize}

\begin{figure}[H]
        \centering
        \begin{subfigure}[h]{0.4\textwidth}
                \centering    \includegraphics[scale=0.6]{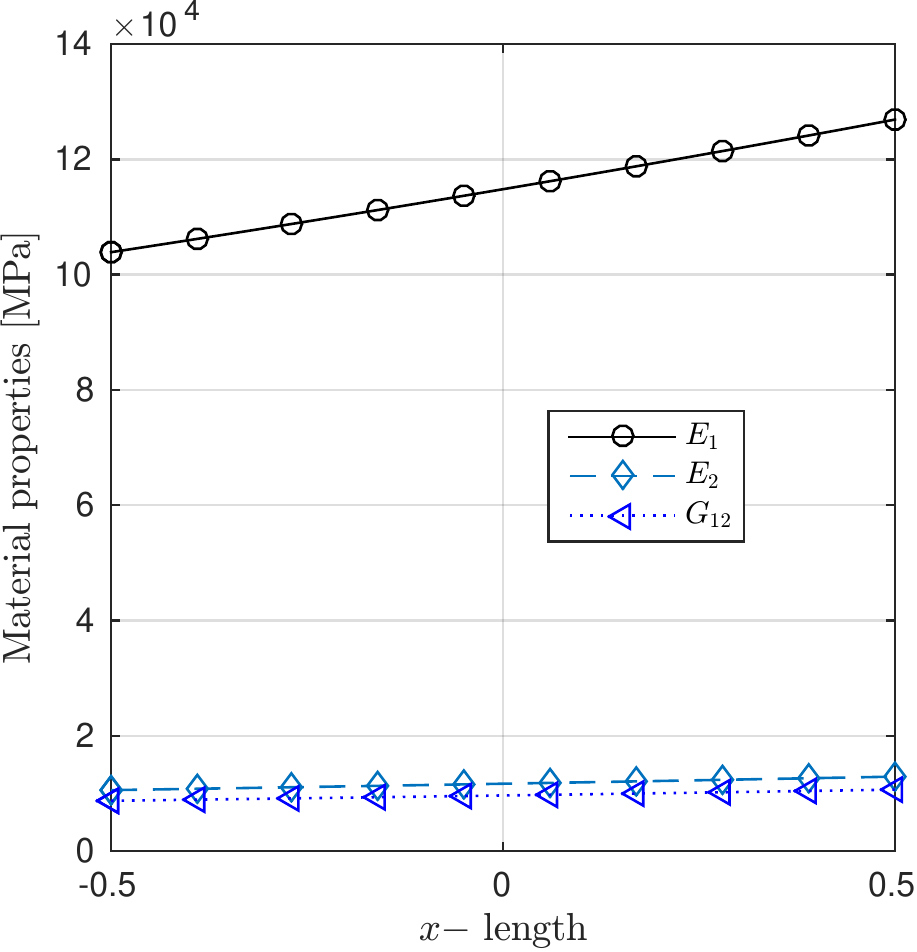}
                \caption{}   \label{fig:crackpathx}
        \end{subfigure}
        \begin{subfigure}[h]{0.4\textwidth}
                \centering    \includegraphics[scale=0.6]{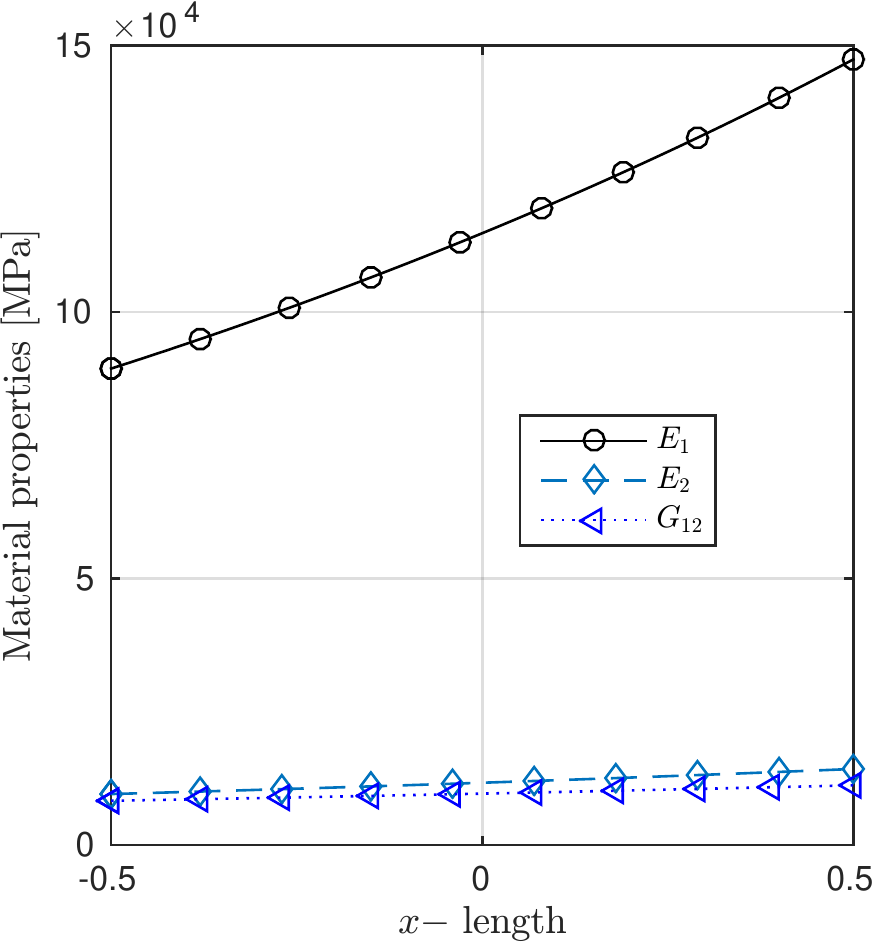}
                \caption{}    \label{fig:crackpathxn}
        \end{subfigure}
        \\
        \begin{subfigure}[h]{0.4\textwidth}
                \centering    \includegraphics[scale=0.6]{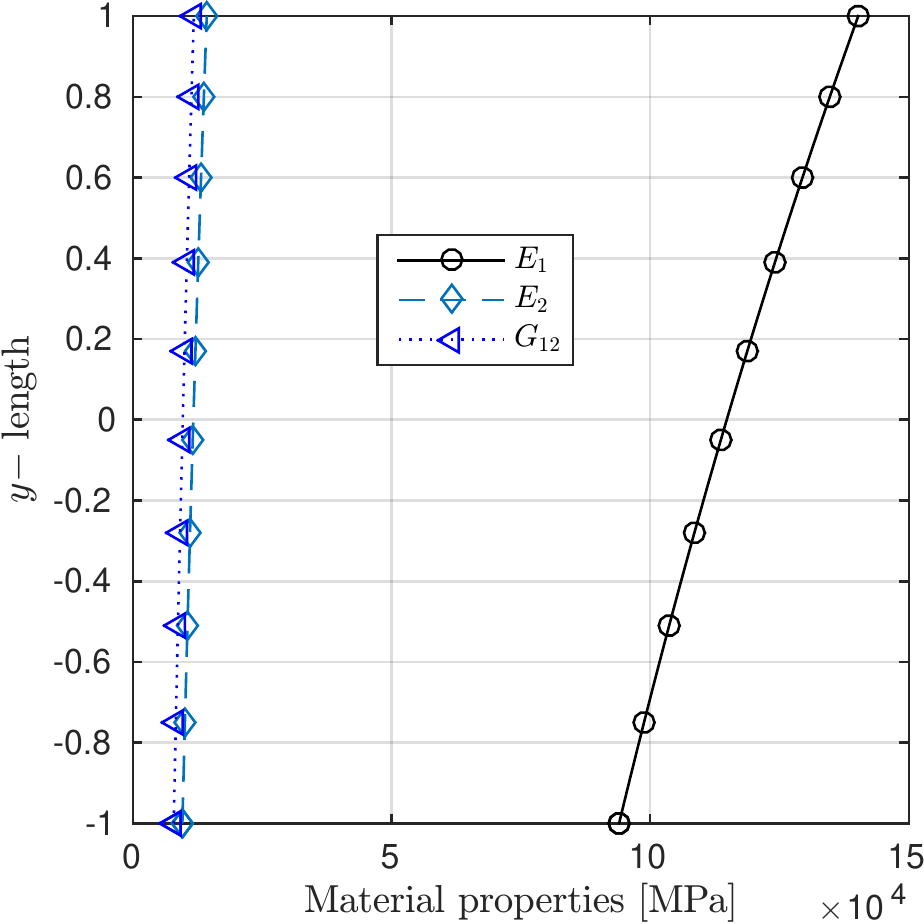}
                \caption{}   \label{fig:crackpathy}
        \end{subfigure}
        \begin{subfigure}[h]{0.4\textwidth}
                \centering   \includegraphics[scale=0.6]{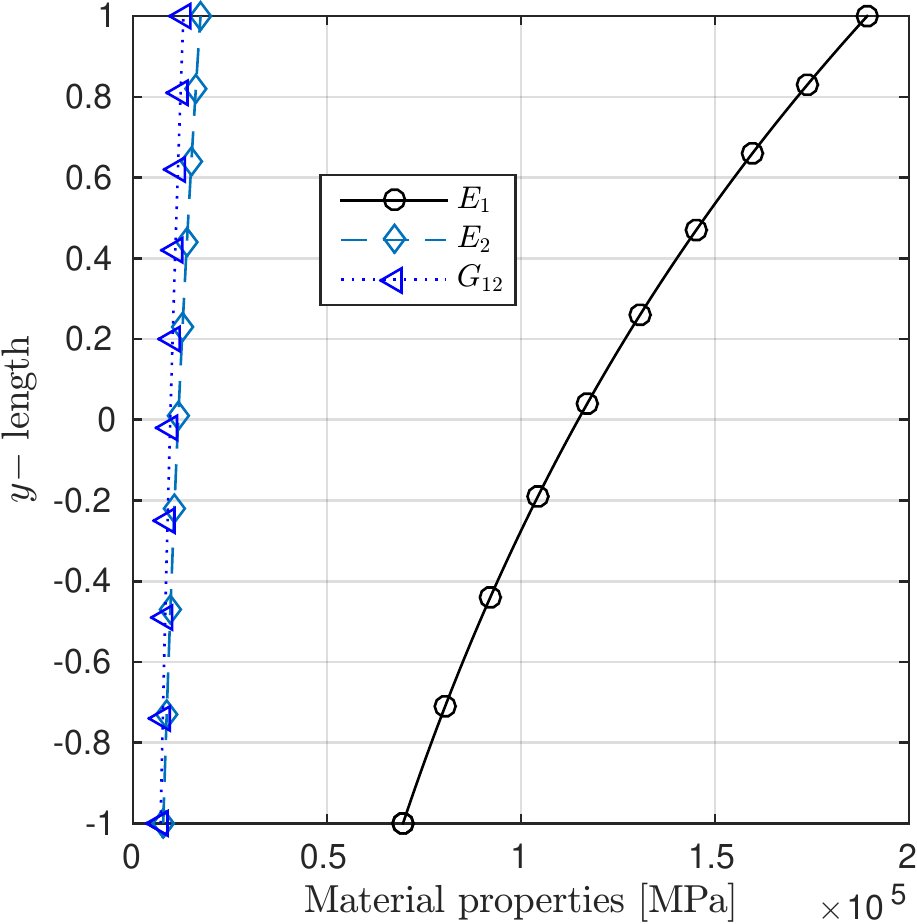}
                \caption{}   \label{fig:crackpathyn}
        \end{subfigure}
        \caption{FGM orthotropic material gradation in (a) $x-$direction with proportional material gradation (b) $x-$direction with non-proportional material gradation (c) $x-$direction with proportional material gradation, and (d) $y-$direction with non-proportional material gradation}
        \label{fig:materialvariation_orthoFGMs}
\end{figure}

For simplicity, we assume that the material property variation follows an exponential gradation, as characterized by the indices $\alpha, \beta, \gamma$ - see Fig. \ref{fig:domian_edge_ortho}. As shown in Fig. \ref{fig:materialvariation_orthoFGMs}, two different scenarios have been considered: (i) proportional and (ii) non-proportional gradation strategies, in terms of the material indices. In the former, the indices, $\alpha,~\beta,~\gamma$ are set to 0.2, whilst for non-proportional variation, we choose, $(\alpha, \beta, \gamma) = (0.5,0.4,0.3)$. The material constants are chosen as: $E_1^0 =$ 114.8 GPa, $E_2^0=$ 11.7 GPa, $G_{12}^0 =$ 9.66 GPa, and the critical toughness $\mathcal{G}_c =$ 2.7 MPa mm. With respect to the critical energy release rate $\mathcal{G}_c$, two cases are considered; one, where no material gradation is assumed and a \emph{graded} one that follows the material gradation depicted in Fig. \ref{fig:domian_edge_ortho}. For all results, the orthotropic material orientation is take to be $\theta=0^\circ$.

\begin{figure}[H]
    \centering
    \begin{subfigure}[h]{0.4\textwidth}
                \centering    \includegraphics[scale=0.7]{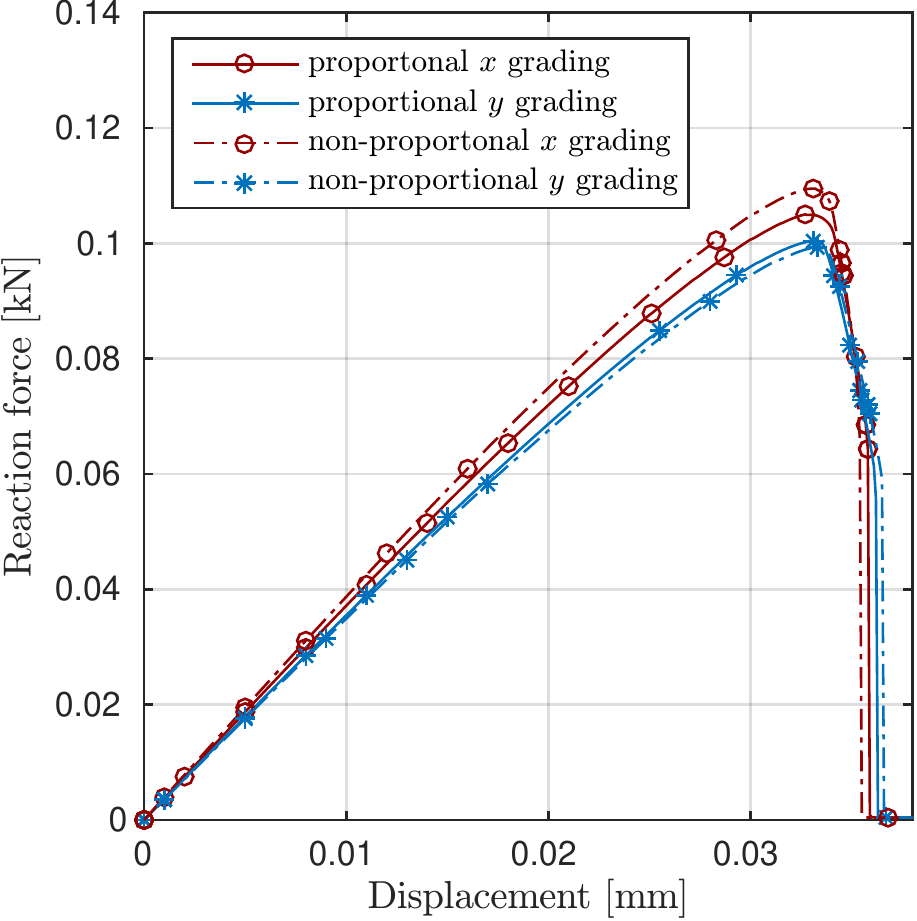}
                \caption{}   \label{fig:crackpathy}
        \end{subfigure}
        \hspace{5mm}
        \begin{subfigure}[h]{0.4\textwidth}
                \centering    \includegraphics[scale=0.7]{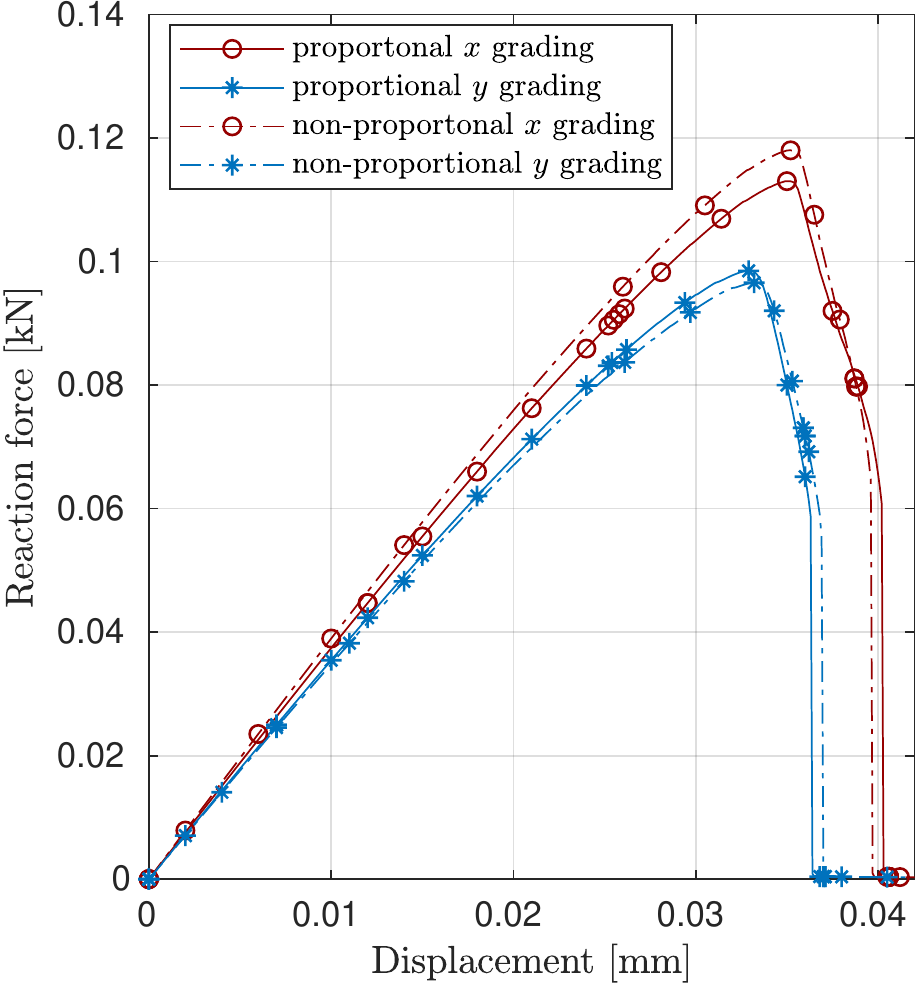}
                \caption{}   \label{fig:crackpathy}
        \end{subfigure}
   
    \caption{Load-displacement response for FGM orthotropic specimen with (a) with constant toughness $\mathcal{G}_c$, and (b) with varying toughness $\mathcal{G}_c (\bm{x})$.}
    \label{fig:my_label}
\end{figure}

In all cases, given that the same value of $\theta$ is considered, the predicted crack trajectories follow an almost identical path. However, differences can be seen in the load-displacement curves, as shown in Fig. \ref{fig:my_label}. Consider first the effect of a proportional or non-proportional material gradation, the same qualitative trends are seen in in both Fig. \ref{fig:my_label}a and Fig. \ref{fig:my_label}b. For the case of material gradation in $x-$direction, the non-proportional material gradation shows stiffer response than the proportional material gradation. This trend is reversed for the case of material gradation in $y-$direction, where differences are minimal. Differences are due to the crack tip non-homogeneity, which affects the mode mixity. Consider now the influence of spatially varying the material critical energy release rate $\mathcal{G}_c$ toughness, i.e. Fig. \ref{fig:my_label}a versus Fig. \ref{fig:my_label}b. It can be seen that differences are substantial in the case of material grading along the $x$-direction. In agreement with expectations, the propagating crack encounters an increasing resistance to fracture as the magnitude of $\mathcal{G}_c$ at the crack tip raises with crack advance.

\section{Conclusions}
\label{Sec:Concluding remarks}

We have presented a novel framework for modelling fracture problems in orthotropic functionally graded materials (FGMs). The framework builds upon the phase field fracture method for FGMs and an adaptive mesh technique based on a recovery based error indicator and quadtree decomposition. Results show the capability of the model in capturing complex crack trajectories, not known \textit{a priori}, while minimising the computational cost. The numerical framework is validated by comparing to experimental and numerical results on non-graded orthotropic materials. A good agreement is observed. Then, calculations are shown for orthotropic FGMs and the role of the material gradation indeces explored. Topics of interest for future work involve extending the present framework to dynamic crack growth, three dimensions problems and enabling mesh coarsening behind the crack.

\section{Acknowledgments}
\label{Acknowledge of funding}

E. Mart\'{\i}nez-Pa\~neda acknowledges financial support from the Royal Commission for the 1851 Exhibition through their Research Fellowship programme (RF496/2018).




\bibliographystyle{elsarticle-num}
\bibliography{mylibrary}

\end{document}